\documentstyle[12pt,subeqn]{article}

\newcommand\vek[1]{\mbox{\rmfamily\bfseries\itshape#1}}

\begin{document}

\title{Charge Correlations in a Coulomb System Along a Plane Wall: a
Relation Between Asymptotic Behavior and Dipole Moment}

\author{
B. Jancovici$^1$ and L. {\v S}amaj$^{1,2}$
}

\maketitle

\begin{abstract}
Classical Coulomb systems at equilibrium, bounded by a plane 
dielectric wall, are studied.
A general two-point charge correlation function is considered.
Valid for any fixed position of one of the points,
a new relation is found between the algebraic tail of the correlation
function along the wall and the dipole moment of that function.
The relation is tested first in the weak-coupling (Debye-H\"uckel) 
limit, and afterwards, for the special case of a plain hard wall, 
on the exactly solvable two-dimensional two-component plasma 
at coupling $\Gamma=2$, and on the two-dimensional one-component 
plasma at an arbitrary even integer $\Gamma$. 
\end{abstract}


\medskip
\noindent LPT Orsay 01-18

\vfill

\noindent $^1$ Laboratoire de Physique Th{\'e}orique, Universit{\'e} de
Paris-Sud, B{\^a}timent 210, 91405 Orsay Cedex, France (Unit{\'e} Mixte
de Recherche no. 8627 - CNRS); 

\noindent e-mail: Bernard.Jancovici@th.u-psud.fr and
Ladislav.Samaj@th.u-psud.fr 

\noindent $^2$ On leave from the Institute of Physics, Slovak Academy of
Sciences, Bratislava, Slovakia

\newpage
 
\renewcommand{\theequation}{1.\arabic{equation}}
\setcounter{equation}{0}

\section{Introduction and Summary}
Near a plane wall impenetrable to the particles (hard wall), the charge
correlations of a classical (i.e. non-quantum) Coulomb system
(plasma, electrolyte, \ldots) at equilibrium have special features 
(see review \cite{Martin}). 
On one hand, they have only an algebraic decay along the wall 
(while the bulk charge correlations decay faster than any inverse 
power law), and their asymptotic form obeys a simple sum rule. 
On the other hand, the charge correlation function
carries a dipole moment (while, in the bulk, there is no such dipole
just for symmetry reasons), and this dipole moment obeys another simple
sum rule. 
Some relation between algebraic tail and dipole moment is 
expected: it is the asymmetry of the screening cloud of a 
particle sitting near the wall which induces that long-range tail in
the charge correlation along the wall. 
In the present paper, we make this relation quantitative.
  
The general classical Coulomb systems under consideration consist of $s$ 
species $\alpha = 1,\ldots,s$ with the corresponding charges $q_{\alpha}$,
plus perhaps a fixed background of density $n_0$ and charge density 
$\rho_0$. Two cases are of particular interest: the one-component plasma
(OCP) which corresponds to $s=1\ (q_1=q),\ \rho_0=-qn_0\ne 0$
and the two-component plasma (TCP) which corresponds to
$s=2\ (q_1=q,q_2=-q),\ \rho_0=n_0=0$. The presence of a solvent is
mimicked by embedding the system in a continuous medium of dielectric
constant $\epsilon$. The walls are made of a material of dielectric
constant $\epsilon_W$. The particles (and the background if any)
interact via the Coulomb potential plus perhaps some short-range forces.
For a $\nu$-dimensional system (in what follows, we will restrict
ourselves to dimensions $\nu = 2, 3$), the Coulomb potential in vacuum 
at position ${\vek r}$, induced by a unit charge at the origin,
defined as the solution of the Poisson equation, is
\begin{equation} \label{1.1}
v({\vek r})  =
\left\{ 
\begin{array}{rl}
-\displaystyle \ln \left( {\vert {\vek r} \vert \over r_0} \right) ,
& \nu = 2 \\ \displaystyle {1\over \vert {\vek r} \vert} \ , 
& \nu = 3 
\end{array}
\right.
\end{equation}
where $r_0$ is a length scale. It should be remembered that, when 
$\nu =2$, the system, with logarithmic interactions, is expected to
mimic some general properties of three-dimensional Coulomb systems, but
does \emph{not} represent ``real'' charged particles confined to a
plane. For two- and more-component plasmas containing pointlike
particles, the singularity of $v({\vek r})$ at the origin prevents the
thermodynamic stability against the collapse of positive-negative
pairs of charges: in two dimensions for small enough temperatures,
in three dimensions for any temperature.
In such cases, the above-mentioned short-range forces (e.g. hard cores) 
are needed, without effect on the results of this paper.

We now define some notations. The microscopic densities of charge
and of particles of species $\alpha$ are given respectively by
\begin{equation} \label{1.2}
{\hat \rho}({\vek r}) = \rho_0+\sum_{\alpha} q_{\alpha}
{\hat n}_{\alpha}({\vek r}), \quad \quad
{\hat n}_{\alpha}({\vek r}) = \sum_i \delta_{\alpha,\alpha_i}
\delta({\vek r}-{\vek r}_i)
\end{equation}
where $i$ indexes the charged particles.
The thermal average at the inverse temperature $\beta = 1/(kT)$
will be denoted by $\langle \ldots \rangle$.
At one-particle level,
\begin{equation} \label{1.3}
\rho({\vek r}) = \langle {\hat \rho}({\vek r})\rangle, \quad \quad
n_{\alpha}({\vek r}) = \langle {\hat n}_{\alpha}({\vek r})\rangle
\end{equation}
At two-particle level, one introduces the two-body densities
\begin{subequations} \label{1.4}
\begin{equation} \label{1.4a}
n_{\alpha \beta}({\vek r},{\vek r}') = \left\langle \sum_{i\ne j} 
\delta_{\alpha ,\alpha_i}\delta_{\beta ,\alpha_j}
\delta({\vek r}-{\vek r}_i) \delta({\vek r}'-{\vek r}_j)
\right\rangle
\end{equation}
and the corresponding Ursell functions
\begin{equation} \label{1.4b}
U_{\alpha \beta}({\vek r},{\vek r}') = 
n_{\alpha \beta}({\vek r},{\vek r}')
-n_{\alpha}({\vek r}) n_{\beta}({\vek r}')
\end{equation}
\end{subequations}
The truncated charge-charge correlation (structure function) reads
\begin{equation} \label{1.5}
S({\vek r},{\vek r}') = \langle {\hat \rho}({\vek r})
{\hat \rho}({\vek r}') \rangle^{{\rm T}} =
\langle {\hat \rho}({\vek r}) {\hat \rho}({\vek r}') \rangle -
\langle {\hat \rho}({\vek r})\rangle 
\langle {\hat \rho}({\vek r}') \rangle
\end{equation}
In the case of the OCP, $S$ takes the form
\begin{equation} \label{1.6}
S({\vek r},{\vek r}') = q^2 \left[
U({\vek r},{\vek r}') + n({\vek r})
\delta({\vek r}-{\vek r}') \right]
\end{equation}
For our purpose, it is useful to introduce ``conditional'' densities.
Let $n_{\alpha}({\vek r}\vert Q,{\vek R})$ be the density of
$\alpha$-particles at point ${\vek r}$ when there is a charge $Q$ 
fixed at ${\vek R}$.
Evidently, if $\beta = 1,\ldots,s$ belongs to the set of charged 
species forming the plasma, it holds
\begin{equation} \label{1.7}
n_{\alpha}({\vek r}\vert q_{\beta},{\vek r}') n_{\beta}({\vek r}')
=  n_{\alpha \beta}({\vek r},{\vek r}')
\end{equation}
The excess charge density at point ${\vek r}$, due to the presence
of the charge $Q$ fixed at ${\vek R}$, then is
\begin{equation} \label{1.8}
\rho^{{\rm ex}}({\vek r}\vert Q,{\vek R}) =\left\{ \sum_{\alpha}
q_{\alpha} \left[ n_{\alpha}({\vek r}\vert Q,{\vek R}) -
n_{\alpha}({\vek r}) \right]\right\}+Q\delta({\vek r}-{\vek R})
\end{equation}
In terms of $\rho^{{\rm ex}}$, $S$ is expressible as follows
\begin{eqnarray} \label{1.9}
S({\vek r},{\vek r}') & = & \sum_{\beta} q_{\beta} n_{\beta}({\vek r}')
\rho^{{\rm ex}}({\vek r}\vert q_{\beta},{\vek r}') \nonumber \\
& = & \sum_{\beta} q_{\beta} n_{\beta}({\vek r})
\rho^{{\rm ex}}({\vek r}'\vert q_{\beta},{\vek r})
\end{eqnarray}

Here, we consider a semi-infinite Coulomb system which
occupies the half-space $x>0$ filled with a medium of 
dielectric constant $\epsilon$; we denote by ${\vek y}$ 
the set of $(\nu - 1)$ coordinates normal to $x$.
The plane at $x=0$ is a hard wall impenetrable to the particles.
It may be charged by a uniform surface charge density $\sigma$.
The half-space $x<0$ is assumed to be filled with a material of
dielectric constant $\epsilon_W$.
As a consequence, a particle of charge $q$ at the point
${\vek r}=(x,{\vek y})$ has an electric image of charge
$[(\epsilon - \epsilon_W)/(\epsilon+\epsilon_W)] q$ at
the point ${\vek r}^*=(-x,{\vek y})$ \cite{Jackson}.
Due to invariance with respect to translations along the wall
and rotations around the $x$ direction,
\begin{equation} \label{1.10}
S({\vek r},{\vek r}') = S(x,x';\vert {\vek y}-{\vek y}' \vert)
= S(x',x;\vert {\vek y}-{\vek y}' \vert)
\end{equation}

The cases $\epsilon_W=\infty$ (ideal conductor wall) and $\epsilon_W=0$
(ideal dielectric wall) are special and will not be considered here. 
For finite $\epsilon_W$, the charge structure factor 
$S(x,x';\vert {\vek y}-{\vek y}' \vert)$ has several general properties:
It obeys a condition of electroneutrality 
\begin{equation} \label{1.11}
\int_0^{\infty} {\rm d}x' \int {\rm d} {\vek y} ~
S(x,x';{\vek y}) = 0
\end{equation}
The Carnie and Chan generalization to nonuniform fluids
of the second-moment Stillinger-Lovett condition \cite{Carnie1}
results for the present geometry in the dipole sum rule
\cite{Blum,Carnie2}
\begin{equation} \label{1.12}
\int_0^{\infty} {\rm d}x \int_0^{\infty} {\rm d}x' \int {\rm d}
{\vek y} ~ x' S(x,x';{\vek y}) = - {\epsilon \over 2 \beta
\pi (\nu -1)}, \quad \quad \nu = 2, 3
\end{equation}
The charge-charge correlations decay slowly along the wall 
\cite{Usenko,Jancovici1}. One expects an asymptotic power-law behavior
\begin{equation} \label{1.13}
S(x,x';{\vek y}) \simeq {f(x,x') \over \vert {\vek y} \vert^{\nu}},
\quad \quad \vert {\vek y} \vert \to \infty
\end{equation}
where $f(x,x')$, which as a function of $x$ or $x'$ has a fast decay
away from the wall, obeys the sum rule \cite{Jancovici2,Jancovici3}
\begin{equation} \label{1.14}
\int_0^{\infty} {\rm d}x \int_0^{\infty} {\rm d}x' f(x,x')
= - {\epsilon_W \over 2 \beta [\pi (\nu-1)]^2}, 
\quad \quad \nu = 2,3
\end{equation}

In this work, we establish a general relation
between the structure function $S$ and its asymptotic
characteristics $f$.
Namely, for any value of $x\ge 0$ it is proven that
\begin{equation} \label{1.15}
\int_0^{\infty} {\rm d}x' \int {\rm d}{\vek y} ~ x'
S(x,x';{\vek y}) = {\epsilon \over \epsilon_W} \pi (\nu-1)
\int_0^{\infty} {\rm d}x' ~ f(x,x'), \quad \quad \nu=2,3
\end{equation}
The lhs of (\ref{1.15}) is a dipole moment (in its expression, due to 
the electroneutrality property (\ref{1.11}), $x'$ can be replaced by
$x'-x$). In contrast with the sum rules (\ref{1.12}) and (\ref{1.14}),
the relation (\ref{1.15}) holds for a given $x$, without integration over
it. However, when both sides of (\ref{1.15}) are integrated over $x$ from  
$0$ to $\infty$, it is seen that the sum rule ({\ref{1.14}) for $f$
is a direct consequence of the dipole sum rule (\ref{1.12}),
and vice versa. 

More generally, it can be assumed that the excess charge density 
(\ref{1.8}) also obeys a condition of electroneutrality similar to 
(\ref{1.11}) and has an asymptotic behavior similar to (1.13)
\begin{equation} \label{1.16}
\rho^{{\rm ex}}({\vek r}\vert Q,{\vek R}) \simeq
{F(x\vert Q,X) \over \vert {\vek y} \vert^{\nu}},
\quad \quad \vert {\vek y} \vert \to \infty
\end{equation}
where we have chosen ${\vek R} = (X,{\vek 0})$, with $F$ as a function
of $x$ or $X$ having a fast decay away from the wall. Under these
assumptions, we derive the relation
\begin{equation} \label{1.17}
\int {\rm d}{\vek r} ~ x \rho^{{\rm ex}}({\vek r}\vert Q,{\vek R})
= {\epsilon \over \epsilon_W}
\pi (\nu - 1) \int_0^{\infty} {\rm d}x ~ F(x\vert Q,X),
\quad \quad \nu=2,3
\end{equation}
valid for an arbitrary $Q$. Here too, in the lhs of (\ref{1.17}), $x$
can be replaced by $x-X$. On account of (\ref{1.9}), this more
general relation immediately leads to (\ref{1.15}) with
$f(x,X) = \sum_{\beta} q_{\beta} n_{\beta}(X) F(x\vert q_{\beta},X)$.

The paper is organized as follows.
Section 2 is devoted to a general derivation of the basic
result (\ref{1.17}).
The validity of this result is checked in the 
Debye-H\"uckel limit $\beta\to 0$ (section 3) and in two dimensions
at the special value of the coupling constant 
$\Gamma = \beta q^2 = 2$, for both the TCP \cite{Cornu} (section 4)
and the OCP \cite{Jancovici1} (section 5) in contact with a plain
hard wall $(\epsilon_W = \epsilon =1)$.
In the case of the two-dimensional OCP, we were able to document
the validity of formula (\ref{1.15}) even for an arbitrary
even integer $\Gamma$ by a non-trivial application of sum rules derived
in ref. \cite{Samaj} using the technique of Grassmann variables.

\renewcommand{\theequation}{2.\arabic{equation}}
\setcounter{equation}{0}

\section{General Derivation}
In this section, the abbreviated notation $\rho^{{\rm ex}}({\vek r})$ 
will be used for the excess charge distribution 
$\rho^{{\rm ex}}({\vek r}|Q,{\vek R})$ defined in (\ref{1.8}), with 
${\vek R}=(X,{\vek 0})$. For simplicity, we first consider the case when
there are no dielectric media $(\epsilon=\epsilon_W=1)$. Our derivation
of (\ref{1.17}) is based on the assumption that there are
good screening properties in the bulk: the charge distribution 
$\rho^{{\rm ex}}({\vek r})$ is localized near the wall and the electric 
potential $\phi({\vek r})$ it creates in the plasma at a macroscopic 
distance from the wall vanishes. This electric potential is
\begin{equation} \label{2.1}
\phi({\vek r})=
\int {\rm d}{\vek r}'~v({\vek r}-{\vek r}')\rho^{{\rm ex}}({\vek r}')
\end{equation}
where $v({\vek r}-{\vek r}')$ is the Coulomb interaction (\ref{1.1}).
When $x$ is large enough for the charge distribution
$\rho^{{\rm ex}}({\vek r})$ to be negligible,
$v({\vek r}-{\vek r}')$ can be expanded in powers of $x'$ (with
the notation ${\vek r}=(x,{\vek y})$, etc\ldots), and one obtains 
\begin{eqnarray} \label{2.2}
\phi({\vek r})=
& & \int {\rm d}{\vek y}'~v(x,{\vek y} - {\vek y}')
\int_0^{\infty}{\rm d}x'~ \rho^{{\rm ex}}(x',{\vek y}')
-\int {\rm d}{\vek y}'\frac{\partial v(x,{\vek y}-{\vek y}')}
{\partial x}\int_0^{\infty}{\rm d}x'~x'
\rho^{{\rm ex}}(x',{\vek y}') \nonumber \\
& & + \ldots 
\end{eqnarray}
where the higher-order terms of the expansion involve higher-order
derivatives of $v$ with respect to $x$.  

We shall now consider the Fourier transform of (\ref{2.2}) with respect
to ${\vek y}$, using the convolution theorem. In (\ref{2.2}), let us
define
\begin{equation} \label{2.3}
\sigma^{{\rm ex}}({\vek y}')=\int_0^{\infty}{\rm d}x'~
\rho^{{\rm ex}}(x',{\vek y}')
\end{equation}
(from a macroscopic point of view, $\sigma^{{\rm ex}}({\vek y}')$
can be regarded as a surface charge density). The total charge 
$\int {\rm d}{\vek y}\ \sigma^{{\rm ex}}({\vek y})$ vanishes, as 
required by the perfect screening of $Q$. Also, it results from 
(\ref{1.16}) that $\sigma^{{\rm ex}}({\vek y})$ has the asymptotic 
behavior
\begin{equation} \label{2.4}
\sigma^{{\rm ex}}({\vek y})\simeq\frac {A}{|{\vek y}|^{\nu}}
\end{equation}
where 
\begin{equation} \label{2.5}
A=\int_0^{\infty}{\rm d}x\ F(x|Q,X)
\end{equation}
Therefore, the Fourier transform of $\sigma^{{\rm ex}}({\vek y})$
with respect to ${\vek y}$ has the small wave number behavior 
\begin{equation} \label{2.6}
\tilde{\sigma}^{{\rm ex}}({\vek l})=\int {\rm d}{\vek y}~
\exp(- {\rm i}{\vek l}\cdot {\vek y})
\sigma^{{\rm ex}}({\vek y})\simeq -(\nu -1)\pi A|{\vek l}|
\end{equation}
The Fourier transform of $v(x,{\vek y})$ is 
$(\nu -1)\pi\exp(-|{\vek l}|x)/|{\vek l}|$ and thus the Fourier
transform of $\partial v/\partial x$ is 
$-(\nu -1)\pi\exp(-|{\vek l}|x)$ and so on for higher-order derivatives
of $v$. Using these transforms in the
convolution theorem gives the Fourier transform $\tilde{\phi}(x,{\vek l})$
of (\ref{2.2}), which, at ${\vek l}={\vek 0}$, is found to be
\begin{equation} \label{2.7}
\tilde{\phi}(x,{\vek 0})=(\nu -1)\pi\left[-(\nu -1)\pi A + P\right]
\end{equation} 
where
\begin{equation} \label{2.8}
P=\int{\rm d}{\vek r}~x\rho^{{\rm ex}}({\vek r})
\end{equation}
Since $x$ has been assumed to be large, $\tilde{\phi}(x,{\vek 0})=0$ and
(\ref{2.7}) results into
\begin{equation} \label{2.9}
-(\nu -1)\pi A + P=0
\end{equation} 
On account of (\ref{2.5}) and (\ref{2.8}), (\ref{2.9}) is (\ref{1.17})
in the special case $(\epsilon=\epsilon_W=1)$. 

The generalization to other values of the dielectric constants is
straightforward. In (\ref{2.2}), one must add to 
$\rho^{{\rm ex}}(x',{\vek y}')$ its electric image \cite{Jackson}
$[(\epsilon -\epsilon_W)/(\epsilon +\epsilon_W)]
\rho^{{\rm ex}}(-x',{\vek y}')$.
This is equivalent to keeping the integration range $(0,\infty)$ 
for $x'$, but multiplying the first integral in (\ref{2.2}) by 
$2\epsilon/(\epsilon +\epsilon_W)$ and the second integral by 
$2\epsilon_W/(\epsilon +\epsilon_W)$. This gives the general form of
(\ref{1.17}). 

\renewcommand{\theequation}{3.\arabic{equation}}
\setcounter{equation}{0}

\section{Debye-H\"uckel Limit}
We check the relation (1.17) for the general Coulomb system defined in
the Introduction, along a plane wall, in the weak coupling limit 
$\beta \to 0$ (Debye-H\"uckel limit). It is convenient to introduce 
the Fourier transform with respect to ${\vek y}$ of the excess charge
density (\ref{1.8})
\begin{equation} \label{3.1}
\tilde{\rho}_Q^{{\rm ex}}(x,X,|{\vek l}|)=\int {\rm d}{\vek y}~
\exp(- {\rm i}{\vek l}\cdot {\vek y})\rho^{{\rm ex}}({\vek r}|Q,{\vek R})  
\end{equation}
One defines a bulk inverse Debye length $\kappa$ by
\begin{equation} \label{3.2}
\kappa^2 =2\pi (\nu -1)\beta (\sum_{\alpha}q_{\alpha}^2 n_{\alpha})/
\epsilon
\end{equation}
where $n_{\alpha}$ is the bulk density of species $\alpha$. A minor
generalization of the calculation in ref. \cite{Jancovici1} gives 
$\tilde{\rho}_Q^{{\rm ex}}$ as a sum of its bulk Debye-H\"uckel form
plus a ``reflected'' term:
\begin{eqnarray} \label{3.3}
& & \tilde{\rho}_Q^{{\rm ex}}(x,X,{\vek l}) 
= -\frac{Q\kappa^2}{2(\kappa^2+ l^2)^{1/2}}
\big\{ \exp \left[ -(\kappa^2+ l^2)^{1/2}|x-X|\: \right] 
\nonumber \\ 
& & +\frac{\epsilon (\kappa^2+ l^2)^{1/2}-\epsilon_W
|{\vek l}|}{\epsilon(\kappa^2+ l^2)^{1/2}+\epsilon_W|{\vek l}|}
\exp \left[ -(\kappa^2+ l^2)^{1/2}(x+X)\: \right]\big\}
+Q\delta (x-X)
\end{eqnarray}

The asymptotic behavior of $\rho^{{\rm ex}}({\vek y})$ is determined 
by the kink of its Fourier transform $\tilde{\rho}^{{\rm ex}}({\vek l})$
at $|{\vek l}|=0$, $Q(\epsilon_W/\epsilon)\exp [-\kappa (x+X)]|{\vek
l}|$ (in the sense of distributions, the inverse Fourier 
transform of $|{\vek l}|$ is $-1/[(\nu -1)\pi |{\vek y}|^{\nu}]\:$ ). 
One does find the asymptotic behavior (\ref{1.16}) with
\begin{equation} \label{3.4}
F(x|Q,X)= -\frac{Q\epsilon_W}{\pi (\nu -1)\epsilon }\exp[-\kappa (x+X)] 
\end{equation}
Therefore
\begin{equation} \label{3.5}
\int_0^{\infty}{\rm d}x~F(x|Q,X)=-\frac{Q\epsilon_W}{\pi (\nu -1)\kappa
\epsilon }\exp(-\kappa X) 
\end{equation}

For finding the dipole moment associated to 
$\rho^{{\rm ex}}({\vek r}|Q,{\vek R})$, one first notes that the
integral of $\rho^{{\rm ex}}$ over ${\vek y}$ is the Fourier transform
(\ref{3.3}) taken at ${\vek l}=0$. The integral over $x$ is easily
computed, and one checks that (\ref{1.17}) is obeyed. 

\renewcommand{\theequation}{4.\arabic{equation}}
\setcounter{equation}{0}

\section{Two-Dimensional Two-Component Plasma}
We now check (\ref{1.17}) on the two-dimensional TCP at the special 
value of the coupling constant $\Gamma=\beta q^2=2$, in the case of
a plain rectilinear hard wall. Without loss of generality, we shall
take the charges $\pm q$ as $\pm 1$. The corresponding correlation 
functions are known \cite{Cornu}. 
We would like to consider the excess charge density (\ref{1.8}) in the
special case $Q=q=1$, i.e. when the particle fixed at ${\vek R}=(X,0)$
is one of the particles (say a positive one) of the system. Although, 
at $\Gamma=2$, for a given fugacity, the densities diverge, the
Ursell functions are finite. Therefore, instead of 
$\rho^{{\rm ex}}({\vek r}|+1,{\vek R})$, we consider the finite quantity
proportional to its non-self part
\begin{equation} \label{4.1}
n_+({\vek R})[\rho^{{\rm ex}}({\vek r}|+1,{\vek R})-Q\delta({\vek r}-
{\vek R})]=
U_{++}({\vek r},{\vek R})-U_{-+}({\vek r},{\vek R}) 
\end{equation}
($U_{s_1 s_2}$ was called $\rho_{s_1 s_2}^{(2)T}$ in ref. \cite{Cornu}).
In that same reference, the possible surface charge density carried by 
the wall was chosen as $-\sigma$ and we shall keep this choice 
in the present section.

The model has a rescaled fugacity $m$ [which has the dimension of an
inverse length such that the bulk correlation length is $1/(2m)$]. 
The Ursell functions in (\ref{4.1}) are expressible in terms of
auxiliary functions $g_{s+}(x,X,y)$, where $s = \pm$, 
as\footnote{We use the symmetry
relations (2.15) of ref. \cite{Cornu}, the first one of which is
misprinted: its rhs should be replaced by its complex conjugate.} 
\begin{equation} \label{4.2}
U_{s+}({\vek r},{\vek R})=-sm^2|g_{s+}(x,X,y)|^2
\end{equation}
The $g$ functions are given by their Fourier transforms
\begin{equation} \label{4.3}
\tilde{g}_{s+}(x,X,l)=\int_{-\infty}^{\infty}{\rm d}y~\exp(-{\rm i}ly)
g_{s+}(x,X,y)
\end{equation}
which are
\begin{subequations} \label{4.4}
\begin{eqnarray}
\tilde{g}_{++}(x,X,l) & = & \frac{m}{2k}\big\{\exp\left( -k|x-X|\,\right) 
-\exp[-k(x+X)]\big\},\ \ l<0   \label{4.4a} \\ 
\tilde{g}_{++}(x,X,l) & = &  \frac{m}{2k}\big\{\exp\left( -k|x-X|\,\right) 
+\frac{k-l+2\pi\sigma}{k+l-2\pi\sigma}\exp[-k(x+X)]\big\},\ l>0 
\label{4.4b} \\
\tilde{g}_{-+}(x,X,l) & = & \frac{1}{2k}\big\{[l-2\pi\sigma +k\:
{\rm sign}(x-X)]\exp\left( -k|x-X|\,\right) \nonumber \\
& & - ( k+l-2\pi\sigma ) \exp[-k(x+X)]\big\},\ \ l<0 \label{4.4c} \\
\tilde{g}_{-+}(x,X,l) & = &  \frac{1}{2k}\big\{[l-2\pi\sigma +k\:
{\rm sign}(x-X)]\exp\left( -k|x-X|\,\right)  \nonumber \\
& & + ( k-l+2\pi\sigma )\exp[-k(x+X)]\big\},\ \ l>0  \label{4.4d}
\end{eqnarray}
\end{subequations}
where $k=[m^2+(l-2\pi\sigma)^2]^{1/2}$.

The asymptotic behavior of the $g(y)$ functions is governed by the 
discontinuity of their Fourier transforms $\tilde{g}(l)$ at $l=0$.
In the sense of distributions, the inverse Fourier transform of 
${\rm sign}(l)$ is ${\rm i}/(\pi y)$. Thus,
\begin{subequations} \label{4.5}
\begin{eqnarray}
g_{++}(x,X,y) & \simeq & \frac{{\rm i}m}{2\pi (k_0-2\pi\sigma)y}
\exp[-k_0 (x+X)] \label{4.5a} \\
g_{-+}(x,X,y) & \simeq & \frac{{\rm i}}{2\pi y}\exp[-k_0 (x+X)] 
\label{4.5b} 
\end{eqnarray}
\end{subequations}
where $k_0=[m^2+(2\pi\sigma)^2]^{1/2}$. Using (\ref{4.5}) in (\ref{4.2})
and (\ref{4.1}) gives
\begin{equation} \label{4.6} 
U_{++}({\vek r},{\vek R})-U_{-+}({\vek r},{\vek R})
\simeq \frac{F(x|X)}{y^2},\ \ |y|\rightarrow \infty
\end{equation}
where
\begin{equation} \label{4.7}
F(x|X)=-\frac{k_0}{2\pi^2}(k_0+2\pi\sigma)\exp[-2k_0(x+X)]
\end{equation}
Therefore
\begin{equation} \label{4.8}
\int_0^{\infty}{\rm d}x~F(x|X)=-\frac{1}{4\pi^2}(k_0+2\pi\sigma)
\exp(-2 k_0 X)
\end{equation}

For computing the dipole moment 
\begin{equation} \label{4.9}
P=\int_0^{\infty}{\rm d}x~(x-X)
\int_{-\infty}^{\infty}{\rm d}y~[U_{++}({\vek r},{\vek R})
-U_{-+}({\vek r},{\vek R})]
\end{equation}
one first considers the integral over $y$. With regard to (\ref{4.2}) 
and (\ref{4.3}),
\begin{equation} \label{4.10}
\int_{-\infty}^{\infty}{\rm d}y~[U_{++}({\vek r},{\vek R})
-U_{-+}({\vek r},{\vek R})]=-\frac{m^2}{2\pi}\int_{-\infty}
^{\infty}{\rm d}l~\left\{[\tilde{g}_{++}(x,X,l)]^2
-[\tilde{g}_{-+}(x,X,l)]^2\right\}
\end{equation}
In the special case $\sigma=0$, using (\ref{4.4}) in (\ref{4.10}) gives
for the dipole moment, after some calculation,
\begin{equation} \label{4.11}
P=-\frac{m^2}{2\pi}\int_0^{\infty}{\rm d}x~(x-X)\int_0^{\infty}
{\rm d}l~\left\{\exp\left( -2k|x-X|\,\right)
+\frac{k-l}{m}\exp[-2k(x+X)]\right\}, \ \sigma=0
\end{equation}
Performing the integral over $x$ before the one over $l$ gives
\begin{equation} \label{4.12}
P=-\frac{m}{4\pi}\exp(-2mX),\ \ \sigma=0
\end{equation}
Comparing (\ref{4.8}) when $\sigma=0$ with (\ref{4.12}), one checks that 
(\ref{1.17}) is obeyed (here $\epsilon=\epsilon_W=1$).

For checking (\ref{1.17}) in the general case $\sigma\neq 0$, it is
enough (and easier) to check that the derivatives with respect to
$2\pi\sigma$ of both its sides are equal. From (\ref{4.8}) one obtains
\begin{equation} \label{4.13}
\frac{{\rm d}}{{\rm d}(2\pi\sigma)}\int_0^{\infty}{\rm d}x~F(x|X)=
-\frac{1}{4\pi^2}\left(1+\frac{2\pi\sigma}{k_0}\right)
\left(1-4\pi\sigma X\right) \exp(-2 k_0 X)
\end{equation} 
On the other hand, the integrand in the rhs of (\ref{4.10}) depends on
$l$ and $\sigma$ only through the combination $l'=l-2\pi\sigma$, and its
analytic form changes at $l'=-2\pi\sigma$ from some function $f_1(l')$
to some other function $f_2(l')$, as apparent in (\ref{4.4}). 
Therefore, the corresponding integral can be rewritten in the form 
\begin{equation} \label{4.14}
I=\int_{-\infty}^{-2\pi\sigma}{\rm d}l'~f_1(l')
+\int_{-2\pi\sigma}^{\infty}{\rm d}l'~f_2(l')
\end{equation}
and one obtains
\begin{equation} \label{4.15}
\frac{{\rm d}I}{{\rm d}(2\pi\sigma)}=-f_1(2\pi\sigma)+f_2(2\pi\sigma)
\end{equation}
This gives d/d$(2\pi\sigma)$ of the integral over $y$ in (\ref{4.9}). 
Then, one computes the integral over $x$, with the result that 
${\rm d}P/{\rm d}(2\pi\sigma)$ is $\pi$ times (\ref{4.13}). This
completes the check of (\ref{1.17}) on the present model.  

\renewcommand{\theequation}{5.\arabic{equation}}
\setcounter{equation}{0}

\section{Two-dimensional one-component plasma}
We consider the two-dimensional OCP in contact with a
plain hard wall $(\epsilon_W = \epsilon = 1)$, localized at $x=0$
and charged by a ``surface'' charge density $q\sigma$, first
at the coupling constant $\Gamma= \beta q^2 = 2$, then
for any even integer $\Gamma$ $(\sigma = 0)$.
We work in units such that $\pi n_0 = 1$, where $n_0$ is
the background density.
We intend to check the relation (\ref{1.15}).

\subsection{$\Gamma = 2$}
The one-body density at a distance $x$ from the wall is given by
(see eq. (2.16) of ref. \cite{Jancovici1})
\begin{equation} \label{5.1}
n(x) = n_0 {2\over \sqrt{\pi}} \int_{-\pi\sigma\sqrt{2}}^{\infty}
{\exp \left[ - (t-x\sqrt{2})^2 \right] \over 1 + \Phi(t)} {\rm d}t 
\end{equation}
where
\begin{equation} \label{5.2}
\Phi(t) = {2\over \sqrt{\pi}} \int_0^{\infty}
\exp (-u^2) {\rm d}u 
\end{equation}
is the error function.
The two-body Ursell function (called $\rho^{(2)}_T$ in eq. (2.18) 
of ref. \cite{Jancovici1}) is
\begin{eqnarray} \label{5.3}
U(x,x';\vert y-y'\vert) & = & - n_0^2 \exp \left[
- (x-x')^2 \right] \nonumber \\
& \times & \left\vert 
{2\over \sqrt{\pi}} \int_{-\pi\sigma\sqrt{2}}^{\infty}
{\exp \left\{ - \left[ t-(x+x')/\sqrt{2} \right]^2 
- {\rm i}t(y-y')\sqrt{2} \right\} \over 1 + \Phi(t)} {\rm d}t
\right\vert^2 
\end{eqnarray}
The structure function $S$ is expressed in terms of $n$ and
$U$ in formula (\ref{1.6}).
The asymptotic $f$-function takes the form
(see eq. (2.21) of ref. \cite{Jancovici1})
\begin{equation} \label{5.4}
f(x,x') = - n_0^2 q^2 {2\over \pi} {\exp \left[ -2(x+\pi\sigma)^2
-2(x'+\pi\sigma)^2 \right] \over 
\left[ 1 + \Phi(-\pi\sigma\sqrt{2}) \right]^2}
\end{equation}

It is easy to show that
\begin{equation} \label{5.5}
{\pi \over q^2} \int_0^{\infty} {\rm d}x' ~ f(x,x') = - ~
{1\over \sqrt{2} \pi^{3/2}} {\exp \left[ - 2 (x+\pi \sigma)^2 \right] 
\over 1 + \Phi(-\pi \sigma\sqrt{2})}
\end{equation}
On the other hand, performing first the integration over 
the $y$-coordinate, one obtains
\begin{eqnarray} \label{5.6}
{1\over q^2} \int_0^{\infty} {\rm d}x' \int_{-\infty}^{\infty} 
{\rm d}y ~ x' S(x,x';y)  =  x {2\over \pi^{3/2}}
\int_{-\pi\sigma\sqrt{2}}^{\infty} {\exp \left[ - (t-x\sqrt{2})^2
\right] \over 1 + \Phi(t)} {\rm d}t & & \nonumber \\
- {8\over \sqrt{2} \pi^2} \exp\left( -2 x^2 \right)
\int_{-\pi\sigma\sqrt{2}}^{\infty} {\exp \left( - 2t^2
+2 \sqrt{2} t x \right) \over [ 1 +\Phi(t) ]^2} I(t) {\rm d}t & &
\end{eqnarray}
where
\begin{eqnarray} \label{5.7}
I(t) & = & \int_0^{\infty} {\rm d}x' ~ x' \exp( - 2 x'^2
+2\sqrt{2} t x' ) \nonumber \\
& = & {1\over 4} + {\sqrt{\pi} \over 4} t \left[ 1 + \Phi(t) \right]
\exp(t^2)
\end{eqnarray}
After simple algebra, one arrives at
\begin{equation} \label{5.8}
{1\over q^2} \int_0^{\infty} {\rm d}x' \int_{-\infty}^{\infty}
{\rm d}y ~ x' S(x,x';y) = J_1 + J_2
\end{equation}
where
\begin{subequations}
\begin{eqnarray}
J_1 & = & {1\over \sqrt{2} \pi^{3/2}} \int_{-\pi\sigma\sqrt{2}}^{\infty} 
{\rm d}t {1\over 1+ \Phi(t)} {\partial \over \partial t} 
\exp \left[ - (t-x\sqrt{2})^2 \right] \label{5.9a} \\
J_2 & = & - {2\over \sqrt{2} \pi^2} \int_{-\pi\sigma\sqrt{2}}^{\infty} 
{\rm d}t {\exp(-t^2) \over [1+\Phi(t)]^2}
\exp \left[ - (t-x\sqrt{2})^2 \right] \label{5.9b}
\end{eqnarray}
\end{subequations}
Using the equality
\begin{equation} \label{5.10}
{\exp(-t^2) \over [1+\Phi(t)]^2} = - {\sqrt{\pi} \over 2}
{\partial \over \partial t} \left[ {1\over 1 + \Phi(t)} \right]
\end{equation}
in $J_2$, the consequent integration per partes implies
\begin{equation} \label{5.11}
J_2 = - {1\over \sqrt{2} \pi^{3/2}} {\exp\left[ - 2 (x+\pi\sigma)^2
\right] \over 1 + \Phi(-\pi\sigma\sqrt{2})} - J_1
\end{equation}
Inserting $J_2$ into (\ref{5.8}) and comparing with (\ref{5.5})
one gets the expected relation (\ref{1.15}) with
$\epsilon_W = \epsilon$ and $\nu = 2$.

\subsection{$\Gamma =$ even integer}
Let the OCP with logarithmic interactions be confined to
a compact two-dimensional domain $V$.
The positively oriented contour enclosing the domain $V$,
denoted by $\partial V$, is defined parametrically as
$x=X(\varphi), y=Y(\varphi); \varphi_0 \le \varphi \le \varphi_1$.
For example, the circle enclosing the disk of radius $R$ 
centered at the origin admits the parametrization
$X(\varphi) = R \cos \varphi, Y(\varphi) = R \sin \varphi;
0\le \varphi \le 2\pi$.
Integrals over the $V$-domain can be expressed in terms of
the $\partial V$-contour integrals according to the rule
\begin{subequations} \label{5.12}
\begin{equation} \label{5.12a}
\int_V \left( {\partial Q \over \partial x} - 
{\partial P \over \partial y} \right) {\rm d}x {\rm d}y
= \int_{\partial V} \left( P {\rm d}x + Q {\rm d} y \right)
\end{equation}
where
\begin{equation} \label{5.12b}
\begin{array}{c}
\displaystyle
\int_{\partial V} P(x,y) {\rm d}x = \int_{\varphi_0}^{\varphi_1}
{\rm d} \varphi ~ P[X(\varphi),Y(\varphi)] X'(\varphi) \\
\displaystyle
\int_{\partial V} Q(x,y) {\rm d}y =  \int_{\varphi_0}^{\varphi_1}
{\rm d} \varphi ~ Q[X(\varphi),Y(\varphi)] Y'(\varphi)
\end{array}
\end{equation}
\end{subequations}
The neutralizing background of uniform density $n_0$ creates
the one-particle potential $-q n_0 v_0({\vek r})$ where
\begin{equation} \label{5.13}
v_0({\vek r}) = \int_V {\rm d}^2 r' ~ v\left(
\vert {\vek r} - {\vek r}' \vert \right)
\end{equation}
The corresponding electric field is $-q n_0 {\vek E}_0({\vek r})$
where
\begin{equation} \label{5.14}
{\vek E}_0({\vek r}) = E_0^x({\vek r}) \hat{\vek x} +
E_0^y({\vek r}) \hat{\vek y} = - \nabla v_0({\vek r})
\end{equation}
with $\hat{\vek x}$ and $\hat{\vek y}$ being unit vectors 
in the $x$ and $y$ directions.
For the half-plane of interest,
\begin{equation} \label{5.15}
v_0({\vek r}) = {\rm const} - \pi x^2; \quad \quad
E_0^x = 2\pi x,~  E_0^y = 0
\end{equation} 
For a disk of radius $R$ centered at the origin ${\vek 0}$,
\begin{equation} \label{5.16}
v_0({\vek r}) = {\rm const} - \pi r^2/2; \quad \quad
E_0^x = \pi x,~ E_0^y = \pi y
\end{equation} 

By mapping the two-dimensional OCP onto a discrete one-dimensional
Grassmann field theory for the coupling constant $\Gamma =$
even integer, two new kinds of sum rules for the structure function
$S$ were established in ref. \cite{Samaj} for an arbitrarily
shaped $V$-domain.
The first sum rule reads (see formulae (61a,b) of ref. \cite{Samaj}
where $U$ is called $n^{{\rm T}}$)
\begin{subequations}
\begin{eqnarray}
- \beta n_0 \int_V {\rm d}^2 r' ~ E_0^x({\vek r}') S({\vek r},{\vek r}')
& = & {\partial n({\vek r}) \over \partial x} + \int_V {\rm d}^2 r' ~
{\partial \over \partial x'} U({\vek r},{\vek r}')
\label{5.17a} \\ 
- \beta n_0 \int_V {\rm d}^2 r' ~ E_0^y({\vek r}') S({\vek r},{\vek r}')
& = & {\partial n({\vek r}) \over \partial y} + \int_V {\rm d}^2 r' ~
{\partial \over \partial y'} U({\vek r},{\vek r}')
\label{5.17b}
\end{eqnarray}
\end{subequations}
With the aid of the prescription (\ref{5.12}), 
the last terms on the rhs of these sum rules can be rewritten as follows
\begin{subequations} \label{5.18}
\begin{eqnarray} 
\int_V {\rm d}^2 r' ~ {\partial \over \partial x'} 
U({\vek r},{\vek r}') & = & \int_{\varphi_0}^{\varphi_1}
{\rm d}\varphi ~ U[{\vek r};(X,Y)] Y' \label{5.18a} \\
\int_V {\rm d}^2 r' ~ {\partial \over \partial y'} 
U({\vek r},{\vek r}') & = & - \int_{\varphi_0}^{\varphi_1}
{\rm d}\varphi ~ U[{\vek r};(X,Y)] X' \label{5.18b}
\end{eqnarray}
\end{subequations}
The second sum rule reads (see formula (45a) of ref. \cite{Samaj})
\begin{eqnarray} \label{5.19}
& & - \beta n_0 \int_V {\rm d}^2 r' \left[ {\vek r}'\cdot
{\vek E}_0({\vek r}') \right] S({\vek r},{\vek r}') \nonumber \\
& & \quad \quad \quad
= 2 n({\vek r}) + {\vek r} \cdot \nabla n({\vek r})
+ \int_{\varphi_0}^{\varphi_1} {\rm d}\varphi ~ 
U[{\vek r};(X,Y)] (X Y' - X' Y) 
\end{eqnarray}

For the half-plane, since the Ursell function
$U({\vek r},{\vek r}')$ goes to zero at large 
$\vert {\vek r}-{\vek r}' \vert$, eq. (\ref{5.17a}) yields
\begin{equation} \label{5.20}
- 2 \pi \beta n_0 \int_0^{\infty} {\rm d}x' \int_{-\infty}^{\infty}
{\rm d}y ~ x' S(x,x';y) = {{\rm d} n(x) \over {\rm d} x}
- \int_{-\infty}^{\infty} {\rm d}y ~ U(x,0;y) 
\end{equation}
This is an OCP generalization of the WLMB equations \cite{Lovett},
\cite{Wertheim} which were originally derived for neutral systems.
It is trivial to obtain the dipole sum rule (\ref{1.12}) by
integrating both sides of (\ref{5.20}) over $x$ from 0 to $\infty$,
then taking into account that $\lim_{x\to\infty} n(x) = n_0$, and
finally considering the electroneutrality condition (\ref{1.11})
at $x=0$.

For a disk of radius $R$, the addition of eq. (\ref{5.17a})
multiplied by $x$ and eq. (\ref{5.17b}) multiplied by $y$,
with the substitutions (\ref{5.18}), leads to 
\begin{equation} \label{5.21}
- \pi \beta n_0 \int_{{\rm disk}}{\rm d}^2 r' ~ 
\left( {\vek r}\cdot {\vek r}' \right) S({\vek r},{\vek r}')
= r {{\rm d} n(r) \over {\rm d} r} + \int_0^{2\pi}
{\rm d}\varphi' ~ \left( {\vek r}\cdot {\vek R}' \right)
U({\vek r},{\vek R}')
\end{equation}
where, in polar coordinates, ${\vek R}' = (R,\varphi')$.
Eq. (\ref{5.19}) takes the form
\begin{equation} \label{5.22}
- \pi \beta n_0 \int_{{\rm disk}} {\rm d}^2 r' ~
\vert {\vek r}' \vert^2 S({\vek r},{\vek r}')
= 2 n(r) + r {{\rm d} n(r) \over {\rm d} r} + 
R^2 \int_0^{2\pi} {\rm d}\varphi' ~ U({\vek r},{\vek R}')
\end{equation}
Since $\vert {\vek r} - {\vek r}' \vert^2 = \vert {\vek r} \vert^2
+ \vert {\vek r}' \vert^2 - 2 {\vek r} \cdot {\vek r}'$,
the combination of relations (\ref{5.21}) and (\ref{5.22}) with
the electroneutrality condition (\ref{1.11}) results in
\begin{eqnarray} \label{5.23}
& & - \pi \beta n_0 \int_{{\rm disk}} {\rm d}^2 r' ~
\vert {\vek r} - {\vek r}' \vert^2 S({\vek r},{\vek r}') \nonumber \\
& & \quad \quad \quad
= 2 n(r) - r {{\rm d} n(r) \over {\rm d} r} - 
\int_0^{2\pi} {\rm d}\varphi' ~ ( 2{\vek r}\cdot {\vek R}' - R^2 )
U({\vek r},{\vek R}')
\end{eqnarray}
Let us move the origin to the disk boundary by introducing $x=R-r$.
Eq. (\ref{5.23}) becomes
\begin{eqnarray} \label{5.24}
& & - \pi \beta n_0 \int_0^{2\pi} {\rm d}\varphi' \int_0^R
{\rm d}x' (R-x') \vert {\vek r} - {\vek r}' \vert^2 
S({\vek r},{\vek r}') \nonumber \\
& = & 2 n(x) + (R-x) {{\rm d} n(x) \over {\rm d} x} -
\int_0^{2\pi} {\rm d}\varphi' \left[ 2 R (R-x) 
\cos(\varphi'-\varphi) - R^2 \right] U({\vek r},{\vek R}')
\end{eqnarray}
We now take the limit $R\to\infty$.
The lhs of (\ref{5.24}) is dominated by the large values
of $\vert {\vek r}-{\vek r}' \vert$.
The asymptotic behaviour (\ref{1.13}) is equivalent to
\begin{equation} \label{5.25}
\vert {\vek r} - {\vek r}' \vert^2 
S({\vek r},{\vek r}') \simeq f(x,x') + \ldots
\end{equation}
where the higher-order terms vanish for large
$\vert {\vek r} - {\vek r}' \vert$.
Substituting $\varphi' \to y = R (\varphi' - \varphi)$ on the rhs
of (\ref{5.24}) and collecting all terms of order $R$,
one finally arrives at the equation
\begin{equation} \label{5.26}
- 2 \pi^2 \beta n_0 \int_0^{\infty} {\rm d}x' f(x,x')
= {{\rm d} n(x) \over {\rm d} x}
- \int_{-\infty}^{\infty} {\rm d}y ~ U(x,0;y) 
\end{equation}
formulated for the half-plane.
This equation is exactly of the form (\ref{5.20}),
with the expected identification (\ref{1.15}) for the case
under consideration $\epsilon = \epsilon_W = 1$ and $\nu = 2$.

\section{Conclusion}
We have derived a general relation between the asymptotic behavior of a
charge correlation function along a plane wall and the dipole moment of
that correlation function, in the general form (\ref{1.17}). In the 
particular case of the charge structure factor, that relation becomes
(\ref{1.15}).

In those exactly solvable models for which the two-body correlations are 
known, finding their asymptotic behavior is often easier than directly
computing the dipole moment. However, in the present paper, we did
compute this dipole moment, for the sake of checking the general relation.

The relation (\ref{1.15}) was first observed in
the special case of a two-dimensional one-component plasma, as one more
application of a mapping onto a Grassmann field theory \cite{Samaj}. 
Afterwards, we realized that the relation could be derived in general
for any Coulomb system.

\section*{Acknowledgments}
Section 2 has been improved thanks to a suggestion of an unknown
referee. The stay of L. \v S. in LPT Orsay is supported by a NATO
fellowship. A partial support by Grant VEGA 2/7174/20 is acknowledged.

\end{document}